\documentclass{revtex4}
\usepackage{graphics}
\usepackage{slashed}
\usepackage[tbtags]{amsmath}

\usepackage[T1]{fontenc}
\usepackage[utf8]{inputenc}

\newcommand{\be}{\begin{equation}}
\newcommand{\ee}{\end{equation}}
\newcommand{\bea}{\begin{eqnarray}}
\newcommand{\eea}{\end{eqnarray}}
\newcommand{\bdm}{\begin{displaymath}}
\newcommand{\edm}{\end{displaymath}}
\newcommand{\baa}{\begin{array}}
\newcommand{\eaa}{\end{array}}
\newcommand{\ds}{\displaystyle}
\newcommand{\kc}{\slashed{k}}
\newcommand{\pc}{\slashed{p}}
\newcommand{\qc}{\slashed{q}}
\begin{document}
\title{Anapole Moment of Leptons in the Minimal Supersymmetric Standard Model}
\author{C. Ayd{\i}n
\thanks{\emph{coskun@ktu.edu.tr:} The author is grateful to Prof.Dr. T.M. Aliev for his careful reading of the manuscript and helpful
suggestions about its revision.}%
}                     
%
%
\affiliation{Department of Physics, Karadeniz Technical University, 61080, Trabzon, Turkey}

\begin{abstract}
Using Feynman-'t Hooft gauge and dimensional regularization, the static parity-violating coupling of the neutrinos and charged leptons to an external electromagnetic field is calculated in the minimal supersymmetry standard model (MSSM). From the obtained result of calculations, we have seen that the main contribution come from chargino particles in MSSM.
\end{abstract}


\maketitle

\section{Introduction}

The Standard Model(SM) is a theory concerning the electromagnetic, weak and strong interactions. The SM of elementary particle is a gauge quantum field theory. \cite{ref01,ref02,ref03,ref04,ref05}. The main point of these theories is that they are renormalizable so that the useful higher-order calculations can be done. One of the major triumples of the unify theory is the calculations to high order in $\alpha$, $\mu$ and $r^2$ and their successful comparison with experiment. Although the SM is a beautiful model with predictive power, few unknowns prevail as well as some problems. For example, one of them is to determine to absolute scale of the neutrino mass or the nature of the neutrino, either it is a Dirac or a Majorana particle. Physics beyond the SM has drawn physicist attention for a long time. One  of the most appealing theories to describe physics at the TeV scale is the minimal supersymmmetric extensions of the SM (MSSM) \cite{ref06,ref07}. It also besides giving a solution to the hierarchy problem, it provides us with a good candidate for cold dark matter(CDM),namely, the lightest neutralino which is a Majorana particle.

It is known that in quantum field theory (QFT), the interaction vertex between a single photon and a fermion can be characterized in terms of four electromagnetic (EM) form factors. In the limit of vanishing momentum transfer between the photon and fermion, the form factors encode the static EM properties of the particle, namly, its charge, magnetic dipole moment, electric dipole moment and anapole moment. After the discovery of violation of parity in the weak interaction Zeldovich showed that there is the static coupling which the called the anapole moment of the particle with a propertional to the Fermi constant $G_F$ \cite{ref08}. As a classical counterpart of this form factor, Zeldovich introduced anapole moment in connection with the global electromagnetic properties of a toroid coil that are impossible to describe within the charge or magnetic dipole moment in spite of explicit axial symmetry of the toroid coil. Finally, in 1974 Dubovik and Cheskov \cite{ref08a} determined the toroid moment in framework of classical electrodynamics. Anapole (that is  some composition of electric dipole) and toroid dipole are indeed quite different in the sense of their nature. For example, the anapole cannot  radiate at all while the toroid coil and its point-like model can radiate. In the late 1960's and early 1970's, Dubovik connected the quantum description of the anapole to classical electrodynamics \cite{ref09,ref10,ref11,ref111,ref12}. The  existence  of an anapole moment has been experimentally verified in nuclei of heavy atoms, in 1997, it was measured experimentally in  the nucleus of Cesium-133  and Ytterbium-174 \cite{ref13}. In nucleus, the spin and the circulating orbital motion of the external nucleus generate the nuclear spin I and an associated effective current. The weak force introduces a small toroidal component to this orbital and spin current, thus generating a parity violating anapole moment \cite{ref14,ref15,ref16,ref17,ref18}. 

A lepton in a P-violating theory has the electromagnetic current given as \cite{ref19,ref20,ref21,ref22,ref23,ref23a}
\be
\baa{r}
\ds <\ell(p_f) | J_\mu | \ell(p_i)>=i e \overline{u}(p_f)\left[ \gamma_\mu F_1(q^2) \right. \ds + \,  \mu(q^\mu/2m) \sigma_{\mu \nu} F_2(q^2) + \, d(q^\mu/2m) \sigma_{\mu \nu} \gamma_5 F_3(q^2) \\\\
 \ds \left. \hspace{1cm} + \, {g^2 a \over 16 \pi^2 M^2_W} \gamma_5(q^2 \gamma_\mu -  {\qc}q_\mu) F_4(q^2) \right] u(p_i) 
 \eaa
\ee
\noindent in terms of charge $e$, anamolous magnetic moment $\mu$, electric dipole moment $d$ and anapole moment $a$ with the associated form factors \mbox{satisfying}
\be
\baa{c}
\ds F_1(q^2) = 1 + {1 \over 6} q^2 <r^2> + \mathcal{O}(q^4) , \quad
F_2(q^2)=1+\mathcal{O}(q^2), \\ \\ \ds F_3(q^2)=1+\mathcal{O}(q^2), \quad F_4(q^2)=1+\mathcal{O}(q^2)
\eaa \ee
\noindent where $<r^2>$ is the charge radius. For the neutrinos, the difference exists only in the term $F_1(q^2)$ as 
\be F_1(q^2) =  {1 \over 6} q^2 <r^2> + \mathcal{O}(q^4).\ee  The lepton can not have an electric dipole moment in the CP-conserving theory.  

These form factors are physical observable when {\mbox{$q^2 \rightarrow 0$}}. The physical significance of the form factors is easy to  understood by considering in the nonrelativistic limit. In this limit, the interaction energy with an external electromagnetic field takes the form
\be
H_{int} = - \mu(\overrightarrow{\sigma} \cdot \overrightarrow{B}) - d(\overrightarrow{\sigma} \cdot \overrightarrow{E}) - a(\overrightarrow{\sigma} \cdot \overrightarrow{j})
\ee
where $\overrightarrow{B}$ and $\overrightarrow{E}$ are the magnetic and electric fields, $\overrightarrow{\sigma}$ is the Pauli spin matrix, and 
\mbox{$\overrightarrow{j}=(\overrightarrow{\nabla} \times \overrightarrow{B} - \dot{\overrightarrow{E}})$} is the electric current density at the point where the particle is situated.

The form factors of the charged lepton and the neutrinos (uncharged leptons) have been calculated in gauge field theories. It is known that the neutrinos are massless that it is Weyl neutrinos in the SM. There are many possible extensions of SM  which give rise massive neutrino. The questions of whether neutrinos have masses and if so, whether they are Dirac or Majorana particles are two of the most important issues in both particle physics and astrophysics. While a Dirac particle has four form factors, a Majorana particle has only one which it is the anapole moment \cite{ref22,ref221,ref23,ref23a,ref24,ref25}. It is known that the neutrino charge radius and the anapole moment are defined separately  and one can interprete arbitrary as a charge radius or as an anapole moment in the SM. This is the correct interpretation of the statement often found in the literature that in the SM $a=-<r^2>/6$. Therefore, the SM values for the neutrino charge radii can be interpreted also as value of the corresponding neutrino anapole moments. But, if it is the massive neutrino, it is not valid. It means that anapole moment and neutrino charge radius are different form factors. According to these results, we have an idea that leptons have an anapole moment form factors.      

In 1980, the anapole moment of the electron was calculated by Dombey and Kennedy \cite{ref26} in the SM but they did not note the gauge dependence. In 1987, H.Czyz et al. \cite{ref28,ref29} showed the gauge dependence of the anapole moment of charged leptons in the context of the SM. In 1991, Musolf and Holstein \cite{ref30} showed the anapole moment and the charge radius of fermionic systems are not well defined and cannot be considered as physically observable. They argued that the anapole moment would dominate any scattering amplitude to which it contributes and could on the basis of their definition, be considered an observable in  the special cases. However, in 1992, Gongora-T and Stuart \cite{ref30a} were pointed that these (anapole moment and charge radius)  quantities are finite and well-defined. In 2017, Kouzakov and Studenikin \cite{ref30b} considered the low-energy elastic neutrino-electron collision by taking into account electromagnetic interactions of massive neutrinos.  In 2017, Whitcomb and Latimer performed a scattering calculations which probes the anapole moment with a spinless charged particle. They showed that, in the non-relativistic limit, the cross sections agrees with a  quantum mechanical computation of the cross section for a spinless current scattered by an infinitesimally thin toroidal solenoid \cite{ref31}. As a result, they have obtained the effect of the anapole term on the scattering. The similar problem for the  neutrinos has been studied by many authors for a long time. Bardeen et al. \cite{refbarden} claimed that the neutrino charge radius is ultraviolent divergent therefore it is not a physical quantity. Dvornikov and Studenikin also supported this idea in their studies \cite{Dvornikov1, Dvornikov2}. However, Lee \cite{reflee1} indicated that including the additional terms from the Z-boson diagrams, it is possible to find a gauge-dependent finite quantity. Also, in the one of the their work \cite{reflee2}, it is shown that it is possible to obtain a gauge-independent  value by considering the additional box diagrams. Lucio et al. \cite{reflucio1, reflucio2} had defined neutrino elektroweak charge radius in the linear gauge. Also additional set of diagrams about neutrino form factors were provided by Degrassi et al. \cite{refdegrassi}. Barnabeu et al. also calculated neutrino charged radius in their studies \cite{refbarnabeu1,refbarnabeu2,refbarnabeu3}. They also supported the idea that the neutrino electroweak charge radius is observable. Also they showed that the neutrinos are expected  to contribute to the total cross section neutrino elastic scattering off electrons, quarks and nuclei and anapole moment contribution to cross sections are similar to those of the neutrinos charge radius. Giunti and Studenikin have analyzed and shortly summarized the neutrino electromagnetic interactions  \cite{ref38}.

Anapole interaction is the least understood type of all possible between fermion and vector bosons. Motivated by this study,  there had been a surge of interest in the study of anapole moment from the astrophysical as well as the particle physics point of view, and  also serves as basis to explain the dark matter \cite{ref32,ref33,ref34,ref35,ref36,ref37}. 

In the SM, the neutrinos are neutral and massless fermions. They only interact  with other particles via weak interactions. In the last several decades, neutrino physics has progressed at a breathtaking pace. If neutrinos has non-zero masses, the left handed components of neutrino fields with definite flavor ($\ell$) can be a superposition of the left handed components of the neutrino fields with definite masses $m_i$. We now know that only three active flavors couple to  the vector bosons. The electroweak eigenstates of the neutrinos are linear combinations of their mass eigenstates. Neutrino oscillation results imply that the flavor neutrino fields $\ds \nu_{\ell L}(x)$  are the mixtures of the left-right handed components of the fields of the neutrinos, with define masses as
\be
\ds \nu_{\ell L}(x)={1 \over 2}(1-\gamma_5)\nu_\ell(x)=\sum_{i=1}^3 U_{\ell i} \nu_{i L}(x),\, \ell=e^- ,\mu^-, \tau^-
\ee
where $U$ is the the unitary PMNS mixing matrix relating the neutrino mass eigenstates to the weak eigenstates and $\nu_\ell(x)$ is the field of neutrino (Majorana or Dirac) with mass $m_i$. Flavor fields $\nu_L(x)$ enter into  the SM charged current($CC$)
\be
\mathcal{L}_I^{CC}(x)=-{g \over 2\sqrt{2}}\sum_{ \ell=e,\mu, \tau }\overline{\nu}_{\ell L}(x) \gamma_\alpha \ell_L(x) W^\alpha(x) + h.c
\ee
and neutral current ($NC$) interactions
\be 
\ds \mathcal{L}_I^{NC}(x)=-{g \over 2\cos\theta_W} \sum_\ell \left[ \overline{\nu}_{\ell L}(x) \gamma_\alpha \nu_{\ell L} + \overline{\ell} \gamma_\alpha(g_V+g_A \gamma_5) \ell \right] Z^\alpha(x)
\ee
is the neutrino $NC$. $W^\alpha(x)$ and $Z^\alpha(x)$ are the fields of  $W$ and $Z$ vector bosons. $g$ is the electroweak interaction constant and  $\theta_W$ is the weak(Weinberg) angle \cite{ref38}. The vector and axial-vector couplings are 
\bdm
g_V \equiv t_{3L}(\ell) - 2 q_\ell \sin^2\theta_W, \quad
g_A \equiv t_{3L}(\ell)
\edm
where $t_{3L}(\ell)=-{1 \over 2}$ is the weak isospin charged leptons $\ell$ ($1/2$ for neutrinos) and $q_\ell$ is the charge of $\ell$ in units of $e$, which is equal to $g \sin \theta_W$, is the positron electric charge. 

The interactions of the lepton and sneutrino with chargino 
(lepton-sneutino-chargino vertices), the charged lepton and the charged slepton with  the neutralino (lepton-slepton-neutralino vertices) and
the neutrino and slepton with chargino are correspondingly given by Lagrangian as \cite{ref39,ref40,ref41,ref42,ref43,ref44}
\be
\ds
\mathcal{L}_I^{SUSY}(x)=\sum_\ell \left\{ \left[ \overline{\ell} (N^L P_L + N^R P_R) \chi^o \tilde{\ell} +  \overline{\ell} (C^L P_L + C^R P_R) \chi \tilde{\nu}_\ell \right] 
\, + \, \overline{\chi} C^{'L}P_L \nu_\ell \widetilde{\ell} \right\} + h.c 
\ee
where 
\bdm \ds P_{L,R}={1 \over 2}(1 \mp \gamma_5),\edm
\bdm  \ds N^L=-{\ds g \over \sqrt{2}}\sum_{A X}\left\{ {m_{\ell i} \over M_W\cos \beta}N_{A3} R_{X i}^\ell + 2N_{A1} \tan \theta_W R_{X i+3}^\ell \right\},\edm
\bdm \ds N^R=-{\ds g \over \sqrt{2}}\sum_{A X}\left\{ [ -N_{A2} - N_{A1}  \tan \theta_W] R_{X i}^\ell + {\ds m_{\ell i} \over M_W\cos \beta}N_{A3} R_{X i+3}^\ell \right\}, \edm
\bdm \ds C^L=g \sum_X {\ds m_{\ell i} \over M_W\cos \beta} V_{A2} R_{X i}^\nu, \edm
\bdm \ds C^R=g \sum_X U_{A1} R_{X i}^{\nu},\edm
\bdm \ds C^{'L}= \sum_X \left(g V_{A1} R^{\ell}_{X i} + {\ds m_{\ell i} \over M_W\cos \beta} V_{A2} R_{X i+3}^\ell  \right)\edm
where $P_{L,R}$ are the project operators.  $N$ and  $U, V$ are the diagonalized matrices of neutralino and chargino mass matrix, respectively. $R$ is the diagonalized matrix of slepton or sneutrino mass matrix. The indices $A$ ($1...4$ for neutralinos; $1,2$ for charginos) and $X$($1...6$ for sleptons; $1,2,3$ for sneutrinos) runs over the dimensions of the respective matrices, whereas $i$ as usual runs over the generations, $m_\ell$ is the mass of the $i^{th}$ charged lepton and rest of the parameters carry the standard definitions.

\section{Calculation}

The basic diagrams which contribute to a parity-violating interactions are shown in Fig. \ref{fig1}  and Fig. \ref{fig2}. The calculation here of the neutrinos anapole moment parallels the calculation by Sakakibara \cite{ref27} of the neutrino charge radius, and Dombey and Kennedy \cite{ref26} of  the electron anapole moment. The simplest gauge to choose to calculate these diagrams is Feynman-'t Hooft gauge, and the calculations are performed using the dimensional regularization produce which preserves gauge in variance. This calculation is done, using above interactions Lagrangians, in framework of the minimal supersymmetric extension of the Standard Model (MSSM) 

We choose the Breit frame in which the leptons have momentum initially ($p-q/2$) and  ($p+q/2$), then we obtain the matrix elements for the MSSM in Fig. \ref{fig1} and Fig. \ref{fig2}, respectively by ignoring the neutrino mass in calculations.

\subsection{Calculations for the neutrinos}

\be  \ds M_i={e g^2 \over 8} \int  {d^{2w}k \over (2\pi)^{2w}} \left\{
\ds \dfrac{\ds \gamma^\alpha(1-\gamma_5)(-{\kc}+{\pc}+{{\qc} \over 2}+m_\ell)}{\ds [(-k+p-{q \over 2})^2-m_\ell^2]} \dfrac{\ds \gamma^\mu(-{\kc}+{\pc}-{{\qc} \over 2}+m_\ell)\gamma_\alpha(1-\gamma_5)}{\ds [(-k+p+{q \over 2})^2-m_\ell^2][k^2-M_W^2]} \right\} 
 \ee
\be \ds M_{ii}={eg^2 \over 8}\int {d^{2w}k \over (2\pi)^{2w}} \left\{  \dfrac{\ds \gamma_\beta(1-\gamma_5)({\kc}+m_\ell)\gamma_\alpha }{\ds [k^2-m_\ell^2][(-k+p-{q \over 2})^2-M_W^2]} \dfrac{V^{\alpha \beta \mu}(-k+p-{q \over 2},-k+p+{q \over 2},q)}{[(-k+p+{q \over 2})^2-M_W^2]} \right\}
\ee
\be \ds M_{iii}={i g \over 2 \cos \theta_W}\gamma^\mu(1-\gamma_5) \dfrac{1}{q^2-M_Z^2} \pi_{\mu \nu}^{Z \gamma}(q^2) \ee
where $\ds \pi_{\mu \nu}^{Z \gamma}$ is the mixing tensor for the $\ds \gamma - Z^0$ mixing diagrams evaluated numerically by Dombey and Kennedy \cite{ref26}. They obtained
\bdm
\widetilde{\pi}_{\mu \nu}^{Z \gamma}(q^2)=(3,83 \cdot 10^{-3})e^2 q^2 g_{\mu \nu} + \mathcal{O}(q^4)
\edm
\be  \ds M_{iv} = e \int {d^{2w}k \over (2\pi)^{2w}} \left\{  \dfrac{\ds (C^{'L} P_L)(-{\kc}+{\pc}-{{\qc} \over 2}+m_\chi)}{\ds (k^2-m^2_{\widetilde{\ell}})[(-k+p-{q \over 2})^2-m_\chi^2]} 
 \dfrac{\gamma^\mu(-{\kc}+{\pc}+{{\qc} \over 2}+m_\chi)(C^{'L^\dag}P_R)}{[(-k+p+{q \over 2})^2-m_\chi^2]}
\right\}
 \ee
\be \ds M_v = 2e \int {d^{2w}k \over (2\pi)^{2w}} \left\{  \dfrac{\ds (p-k)^\mu(C^{'L} P_L)}{\ds (k^2-m^2_{\chi})[(-k+p-{q \over 2})^2-m_{\widetilde{\ell}}^2]} 
\dfrac{({\kc}+m_{\chi})(C^{'L^\dag}P_R)}{[(-k+p+{q \over 2})^2-m_{\widetilde{\ell}}^2]}
\right\}
 \ee

Projecting out the axial parts propertional to
\bdm
\dfrac{i e}{16\pi^2}\gamma^\mu \gamma_5 \dfrac{g^2}{M^2_W}q^2
\edm
and after long but straightforward calculations, neglecting the terms of order $(m_\ell/M_W)^2$, we obtain for the anapole part the following expressions as
\be
a_i^\nu={1 \over 6} \ln\left({M_W^2 \over m_\ell^2}\right)-{13 \over 72}
\ee
\be
a_{ii}^\nu=-{7 \over 144}
\ee
\be
a_{iii}^\nu=\left({\cos \theta_W \sin \theta_W\over 4}\right) \, 3,83 \cdot 10^{-3}
\ee
\be
a_{iv}^\nu={1 \over 6}\left({M_W^2 \over m_{\widetilde{\ell}^2}}\right)\left[-{7 \over 6}+2\ln\left({m_{\widetilde{\ell}} \over m_{\chi^-}}\right)\right]\left[{1 \over g^2}(|C^{'L}|^2)\right]
\ee
\be
a_v^\nu={1 \over 6}\left({M_W^2 \over m_{\widetilde{\chi}}^2}\right)\left[-{1 \over 6g^2}(|C^{'L}|^2)\right]
\ee
corresponding to Fig. \ref{fig1}((i)--(v)). Then, the total anapole moment of the neutrinos is
\be
a_T^\nu=a_i^\nu + a_{ii}^\nu + a_{iii}^\nu + a_{iv}^\nu + a_v^\nu.
\ee

\subsection{Calculations for charged leptons}
\be \ds M_i={e^3 \over 4} \int {d^{2w}k \over (2\pi)^{2w}}\dfrac{\ds \gamma^\alpha(g_V+g_A\gamma_5)(-{\kc}+{\pc}+{{\qc} \over 2}+m_\ell)\gamma^\mu(-{\kc}+{\pc}-{{\qc} \over 2}+m_\ell)\gamma_\alpha(g_V+g_A\gamma_5)}{\ds [(-k+p-{q \over 2})^2-m_\ell^2][(-k+p+{q \over 2})^2-m_\ell^2][k^2-M_Z^2]}
\ee
\be \ds M_{ii}={eg^2 \over 8}\int {d^{2w}k \over (2\pi)^{2w}} \dfrac{\ds \gamma_\beta(1-\gamma_5)({\kc}-m_\nu)\gamma_\alpha V^{\alpha \beta \mu}(-k+p-{q \over 2},-k+p+{q \over 2},q)}{\ds [k^2-m_\nu^2][(-k+p-{q \over 2})^2-M_W^2][(-k+p+{q \over 2})^2-M_W^2]}
\ee
\be \ds M_{iii}={i g \over 2 \cos \theta_W}\gamma^\mu(g_V+g_A\gamma_5) \dfrac{1}{q^2-M_Z^2} \pi_{\mu \nu}^{Z \gamma}(q^2) .\ee
\be \ds M_{iv} = e \int {d^{2w}k \over (2\pi)^{2w}} \dfrac{\ds (C^L P_L + C^R P_R)(-{\kc}+{\pc}-{{\qc} \over 2}+m_\chi)\gamma^\mu(-{\kc}+{\pc}+{{\qc} \over 2}+m_\chi)(C^{L^\dag}P_R+C^{R^\dag}P_L)}{\ds (k^2-m^2_{\widetilde{\nu_\ell}})[(-k+p-{q \over 2})^2-m_\chi^2][(-k+p+{q \over 2})^2-m_\chi^2]}
\ee
\be \ds M_v = 2e \int {d^{2w}k \over (2\pi)^{2w}} \dfrac{\ds (p-k)^\mu(N^L P_L + N^R P_R)({\kc}+m_{\chi^0})(N^{L^\dag}P_R+N^{R^\dag}P_L)}{\ds (k^2-m^2_{\chi^0})[(-k+p-{q \over 2})^2-m_{\widetilde{\ell}}^2][(-k+p+{q \over 2})^2-m_{\widetilde{\ell}}^2]}
\ee

Similar to neutrinos, we obtain for the anapole part the following expressions as
\be
a_i^\ell={1 \over 3}(1-4\sin^2\theta_W)\left[\ln\left({M_Z^2 \over m_\ell^2}\right)-{7 \over 12}\right]
\ee
\be
a_{ii}^\ell=-{7 \over 72}
\ee
\be
a_{iii}^\ell= 0.15 \cos \theta_W \sin \theta_W 
\ee
\be
\ds
a_{iv}^\ell={1 \over 12}\left({M_W^2 \over m^2}\right)\left[{m_{\widetilde{\nu}}^2+m_{\chi^-}^2 \over m^2 }+
{m_{\widetilde{\nu}}^2 \over m_{\widetilde{\nu}}^2-m_{\chi^-}^2 } \ln\left({m_{\widetilde{\nu}}^2 \over m_{\chi^-}^2}\right)\right] \cdot \left[{1 \over g^2}(|C^L|^2-|C^R|^2)\right]
\ee
\be
\ds a_v^\ell={1 \over 4}\left({M_W^2 \over m_\ell^2}\right) \left[ - {1 \over 2} 
+ {m_{\chi^0}^6 \over (m_{\chi^0}^2-m_{\widetilde{\ell}}^2)^4} \ln({m_{\widetilde{\ell}}^2 \over m_{\chi^0}^2 }) \right] \cdot \left[{1 \over g^2}(|N^L|^2-|N^R|^2)\right]
\ee
corresponding to Fig. \ref{fig2}((i)--(v)). Then, the total anapole moment of the charged leptons is
\be
a_T^\ell=a_i^\ell + a_{ii}^\ell + a_{iii}^\ell + a_{iv}^\ell + a_v^\ell.
\ee

\section{Conclusion}

The anapole form factor is the most mysterious and ambiguous among the form factors, and the anapole interaction is the least understood type of all possible between fermion and vector bosons. Dubovik and Kuznetsov \cite{ref111}  showed that  the toroid dipole moment is a more convenient electromagnetic characteristic of the particle than the anapole. And they obtained the connection between the anapole moment and  the toroid dipole moment for Majorona neutrinos
\be
A(q^2)=T(q^2)+{ m_i^2-m_f^2 \over q^2-\Delta m^2} \left[ D(q^2) -D(\Delta m^2) \right]
\ee
where the masses of the initial $m_i$ and final $m_f$ lepton eigenstates are equal to each other. Thus, the anapole and the toroidal parametrizations coincide in the case when the current is diagonal on lepton initial and final masses.

\begin{figure}[!h]
\resizebox{1.0\textwidth}{!}{%
  \includegraphics{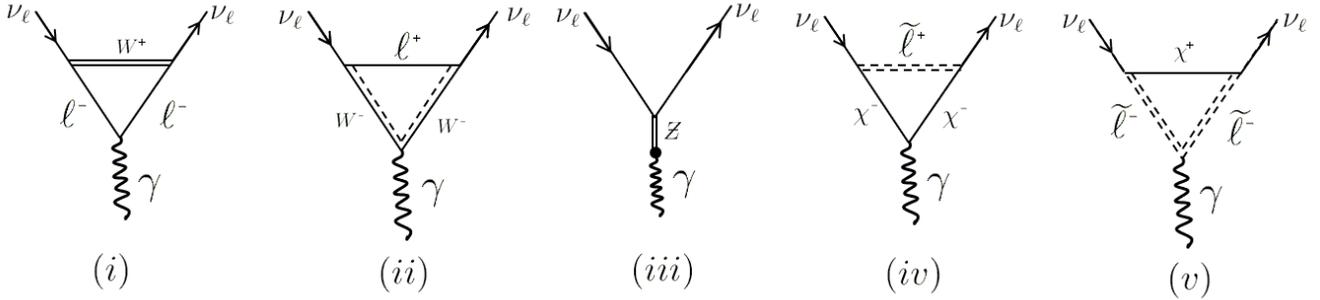}
}
\caption{Basic diagram of the anapole moment in the minimal supersymmetry standard model for the neutrinos}
\label{fig1}       
\end{figure}

\begin{figure}[!h]
\resizebox{1.0\textwidth}{!}{%
  \includegraphics{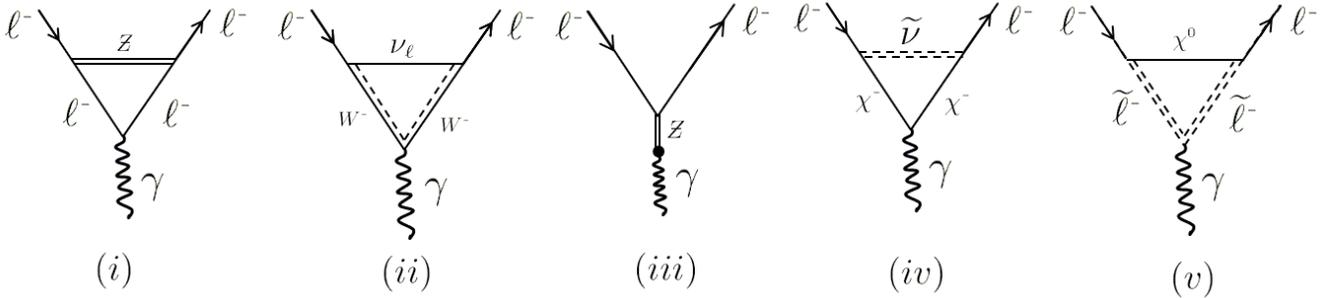}
}
\caption{Basic diagram of the anapole moment in the minimal supersymmetry standard model for the charged leptons}
\label{fig2}       
\end{figure}

 We have computed the anapole moment arising from the exchange of $Z(W)$ and the leptons in the loop, as well as from the exchange of charginos (neutralino) and sleptons in the MSSM \cite{ref27a}. As much as we know, there is no any study in literature for diagrams $(iv)$ and $(v)$ for the leptons (see Figure \ref{fig1}  and Figure \ref{fig2}) for contribution of the anapole moment (for the case of diagrams $(i)$ - $(iii)$ see \cite{ref22}).  Even though this model is perhaps an unrealistically model, it is one of extending of the SM.  Constraints that can be obtained for  this interaction from the entire body of available experiment data are investigated. We think that the most stringent constraints follow from experiments with polarized electrons.Because low energy scattering of polarized electron on electrons and other targets are very sensitive to parity violating effects that are proportional to $(1-4 \sin^2\theta_W)$. However for the neutrinos it is unit.  

 The input parameters must be such as to satisfy a number of experimental constraints \cite{ref45,ref46}. The proper understanding of the nature of cold dark matter from experimental data require the precise analysis of all information that will become available from collider experiments, low-energy experiments astrophysical and cosmological observations \cite{ref47}. We also think that  the anapole interaction may be important for the nature of cold dark matter.

%

These results are very sensitive in the experiment point of view.
Even very small deflection from SM will unambiguously indication of new physics beyond the SM. 
Without performing calculations, we can't say nothing about order of magnitude of considered contributions.
Conducting at present time and planning in future very sensitive experiments on lepton magnetic, electric dipole and anapole moments are one of main research direction in low energy experiments. For this reason presented in this work calculations have direct significance for experiment.

It is well known that there are many scenario in SUYS. However, the large $\tan \beta$ scenario is an interesting issue that has been received attentions in MSSM to find the order of the anapole moment in our calculations. In these scenario, we obtain anapoles for both neutrinos and charged leptons as
\be
\baa{l}
\ds 
a_T^\nu={1 \over 6} \ln \left( {M_W^2 \over m_\ell^2} \right) - {33 \over 144} + { m_\ell^2 \over 48 }
\\ \\ \ds \hspace{0.5cm} \left\{  \left[- {3 m_{\widetilde{\ell}}^2 \over 2(m_{\widetilde{\ell}}^2-m_{\chi^-}^2)} + { m_{\widetilde{\ell}}^4 \over (m_{\widetilde{\ell}}^2-m_{\chi^-}^2)^3} \right] \left[ { 9 m_{\widetilde{\ell}}^4 \over (m_{\widetilde{\ell}}^2-m_{\chi^-}^2)^3} + { m_{\widetilde{\ell}}^6 \over (m_{\widetilde{\ell}}^2-m_{\chi^-}^2)^4} \right] \ln\left({m_{\widetilde{\ell}^2} \over m_{\chi^-}^2}\right) \right\}  \tan^2 \beta
\eaa
\ee
and
\be
a_T^\ell = {1 \over 3}(1-4\sin^2\theta_W)\left[\ln\left({\ds M_Z^2 \over m_\ell^2}\right)-{7 \over 12}\right]-{7 \over 72}+{1 \over 48}
\left[{m_{\chi^-}^2 + m_{\widetilde{\nu}}^2 \over m_\ell^2 } - {m_\ell^2 m_{\widetilde{\nu}}^2 \over (m_{\widetilde{\nu}}^2-m_{\chi^-}^2)^2 }  
 \ln\left({\ds   m_{\chi^-}^2 \over m_{\widetilde{\nu}}^2} \right)\right] \tan^2 \beta .
\ee

It is unnecessary to obtain results for the different values of the parameters. This means that the terms come from SUSY are strongly effective due to the properties of  neutralino and chargino.

The numerical values of the anapole moment for three different Dirac neutrinos and charged leptons are obtained as
\bdm
{g^2 a \over 16 \pi^2 M^2_W} \simeq \left\{ \baa{lclclcl} \ds  6,1 \cdot 10^{-34} cm^2 & \textrm{for} & \nu_e ,  & \quad  & 5,9 \cdot 10^{-23} cm^2 & \textrm{for} & e^{-},
\\ \ds  3,2 \cdot 10^{-34} cm^2 & \textrm{for} & \nu_e ,  & \quad &  1,5 \cdot 10^{-28} cm^2 &  \textrm{for}  & \mu^{-} ,\\
\ds  2,2 \cdot 10^{-34} cm^2 & \textrm{for} & \nu_e , & \quad & 4,9 \cdot 10^{-31} cm^2 &  \textrm{for}  & \tau^{-} 
\eaa
\right. 
\edm
using the values given in Table \ref{tab:1}. We multiply the results by a factor of 2 for Majorona  neutrinos because of the contributions of particle and antiparticle current are equal to each other. It is seen that the contributions of anapole moment for the charged leptons in MSSM is extensively larger than in SM. 

\begin{table}[!h] 
\caption{The particle masses values used in the calculations \cite{ref48}} \label{tab:1}
\centering
\begin{tabular}{|llll|} \hline
{\bf Particle} & {\bf Mass} [GeV]  \quad \quad  \quad \quad \quad  \quad & {\bf Particle} & {\bf Mass} [GeV] \\ \hline & & & \\  
$e^-$ & $5,11 \cdot 10^{-4}$ & $\widetilde{\chi_1^0}$ & $\ge 46$ \\ 
$\mu^-$ & $0,105$ & $\widetilde{\chi_1^\pm}$ & $\ge 94$\\
$\tau^-$ & $1.777$ & $\widetilde{e}$ &  $\ge 107$ \\
$Z^0$ & $91,18$ & $\widetilde{\mu}$ & $\ge 94$ \\
$W^\pm$ & $80.39$ & $\widetilde{\tau}$ &  $\ge 81,9$ \\
$H$ & $125$ & $\widetilde{\nu}$ & $\ge 41$ \\
\multicolumn{2}{|l}{\bf \emph{Some constant values}}& &\\
$G_F$ & = $1,16 \cdot 10^{-5}$ & $1/\alpha$ & =  $137$ \\
$\sin^2 \theta_W$ & = $0.229$ & $\tan \beta$ & = $20$  \\ \hline
\end{tabular}
\end{table}

%
%

\end{document}